\documentclass[showpacs,twocolumn]{revtex4}%
\usepackage{amsfonts}
\usepackage{amsmath}
\usepackage{amssymb}
\usepackage{graphicx}
\usepackage{mathpazo}%
\providecommand{\U}[1]{\protect\rule{.1in}{.1in}}

\begin{document}
\title{Non-Abelian spin-orbit gauge: Persistent spin helix and quantum square ring}
\author{Son-Hsien Chen}
\email{d92222006@ntu.edu.tw}
\affiliation{Department of Physics, National Taiwan University, Taipei 10617, Taiwan}
\author{Ching-Ray Chang}
\affiliation{Department of Physics, National Taiwan University, Taipei 10617, Taiwan}


\begin{abstract}
We re-express the Rashba and Dresselhaus interactions as non-Abelian
spin-orbit gauges and provide a new perspective in understanding the
persistent spin helix [Phys. Rev. Lett. \textbf{97}, 236601 (2006)]. A
spin-orbit interacting system can be transformed into a free electron gas in
the equal-strength Rashba-Dresselhaus [001] linear model, the Dresselhaus
[110] linear model, and a one-dimensional system. A general tight-binding
Hamiltonian for non-uniform spin-orbit interactions and hoppings along
arbitrary directions, within the framework of finite difference method, is
obtained. As an application based on this Hamiltonian, a quantum square ring
in contact with two ideal leads is found to exhibit four states, insulating,
spin-filtering, spin-flipping, and spin-keeping states.

\end{abstract}
\maketitle

\section{Introduction}

The spin-orbit (SO) interaction interests scientists not only for its wide
applications in spintronics devices, but also for its profound fundamental
spin physics. As frequently discussed, the Rashba spin-orbit (RSO) interaction
and Dresselhaus spin-orbit (DSO) interaction exist in two-dimensional electron
gas (2DEG) made of semiconductor heterostructures. The former, with strength
adjustable via the gate voltage\cite{Nitta,Das} results from the inversion
asymmetry of the structure\cite{Rashba}, while the latter is due to lack of
bulk inversion symmetry.\cite{Dresselhaus,Lommer} The SO coupling acts as an
effective magnetic field\cite{Effmag} and thus rotates spin. In general, this
field depends on the electron momentum, while in the two cases,
Rashba-Dresselhaus [001] (linear) model of equal-strength RSO and DSO
couplings and the Dresselhaus [110] (linear) model, it is independent of the
direction of momentum. Therefore, in such two special cases, the precession
angle depends only on the traveling distance but not on the moving-direction
of the electron. Accordingly, the electron precesses as a helix and the
precession pattern, i.e. the spin configuration, is persistent against any
momentum-dependent (but spin-independent) scatterings. The so-called
persistent spin helix\cite{PSHSCZ, PSHHaoger} (PSH) is thus resulted.

Earlier study by Hatano \textit{et. al.}\cite{Non-Abelian} shows that SO
interactions can be regarded as non-Abelian SO or \textit{SU}(2) gauges
$\mathbf{A}^{\text{SO}}=(A_{x}^{\text{SO}},A_{y}^{\text{SO}},0)$ which impose
spin-dependent phases on the traveling electron. By further adjusting the
strength of the RSO coupling and of the magnetic field and neglecting the DSO
coupling, they achieve a perfect spin-filtering ring in which one spin
component gains destructive interference while the other gains constructive
one. Nevertheless, the more realistic case of the coexistence of RSO and DSO
interactions has not been considered.

In this paper, we consider the RSO- and DSO-interacting system subject to an
external magnetic field. We will point out that, even with the coexistence of
RSO and DSO, the SO interactions and magnetic field can as well all be
regarded as gauges. To obtain the discrete (in space) tight-binding (TB)
model, we employ the finite-difference\cite{Databook} (FD) method and
analogize the \textit{SU}(2) gauge with \textit{U}(1) gauge, i.e., analogize
the SO gauge with the magnetic gauge. The condition of performing this analogy
will be specified. We justify this SO-interacting TB model by checking its
consistency with a previously proposed one.\cite{Nicolic} Utilizing these
gauges, we show that: (i) For continuos Hamiltonian, the predicted
PSH\cite{PSHSCZ, PSHHaoger} can be understood easily from the perspective of
gauge transformation. (ii) For discrete TB Hamiltonian, a square ring
functions as a versatile device with four states, insulating,
spin-filtering,\cite{Non-Abelian} spin-flipping and spin-keeping states.

This paper is organized as follows. Both of Secs. \ref{SH} and \ref{SA} are
divided into two parts which focus respectively on the continuos and discrete
cases. The Hamiltonian is studied for the continuos one in Sec. \ref{HCon} and
the discrete one in Sec. \ref{Hdis}. Applications are also given in Secs.
\ref{Acon} and \ref{Adis} for both cases. The Rashba-Dresselhaus [001] linear
Hamiltonian is focused throughout this article, except in Sec. \ref{Adis}
where we consider also the Dresselhaus [110] linear model to demonstrate the
PSH. We summarize in Sec. \ref{Summary}.

\section{Spin-orbit gauge in the Rashba-Dresselhaus [001] linear
Hamiltonian\label{SH}}

We study, in this section, the single-particle Rashba-Dresselhaus [001] linear
Hamiltonian. In Sec. \ref{HCon}, we introduce the RSO and DSO interactions and
show that they can be expressed as gauges. To make this Hamiltonian
numerically treatable, we discretize it in Sec. \ref{Hdis} by considering a TB
model for the free-electron system subject to an external magnetic field.

\subsection{Continuous case\label{HCon}}

Spin-orbit interactions in the 2DEG ($x$-$y$ plane) are commonly modeled by
taking into account the lowest order (linear) in momentum $\mathbf{p}$.
Consider a heterostructure grown along the [001] direction. The corresponding
single-particle Hamiltonian, together with the kinetic energy and the magnetic
gauge $\mathbf{A}^{B}$, reads%
\begin{equation}
H=\frac{\mathbf{\Pi}^{2}}{2m}+\frac{\alpha}{\hbar}\left(  \Pi_{y}\sigma
_{x}-\Pi_{x}\sigma_{y}\right)  +\frac{\beta}{\hbar}\left(  \Pi_{x}\sigma
_{x}-\Pi_{y}\sigma_{y}\right)  \text{,} \label{RDha}%
\end{equation}
with $\alpha$ and $\beta$ denoting the RSO- and DSO- coupling strengths,
respectively. The magnetic field $\mathbf{B}$ is introduced by the kinetic
momentum $\mathbf{\Pi=p-}e\mathbf{A}^{B}\mathbf{/}c$. Since Eq. (\ref{RDha})
has the highest order being quadratic in $\mathbf{\Pi}$\textbf{, }one can
always properly shift the operator $\mathbf{\Pi}$ so that the linear term
disappears. By doing so, the SO gauge,%
\begin{equation}
\mathbf{A}^{\text{SO}}=\left(  A_{x},A_{y}\right)  \equiv\frac{mc}{e\hbar
}\left(  \alpha\sigma_{y}-\beta\sigma_{x},-\alpha\sigma_{x}+\beta\sigma
_{y}\right)  \text{,} \label{SOgauge}%
\end{equation}
due to the presence of SO interactions, and magnetic gauge $A^{B}$ can be
treated on an unified ground. The Hamiltonian Eq. (\ref{RDha}) thus becomes%
\begin{align}
H  &  =\frac{1}{2m}\left(  \mathbf{\Pi-}\frac{e}{c}\mathbf{A}^{\text{SO}%
}\right)  ^{2}-VI_{s}\label{RDgauge}\\
&  =\frac{1}{2m}\left(  \mathbf{p-}\frac{e}{c}\mathbf{A}\right)  ^{2}%
-VI_{s}\text{,}\nonumber
\end{align}
with the constant potential $V=(m/\hbar^{2})\left(  \alpha^{2}+\beta
^{2}\right)  $. The algebra of Pauli matrices $\sigma_{i}^{2}=1$ and $\left\{
\sigma_{i},\sigma_{j}\right\}  =2\delta_{ij}$ with $i,j$ $\in\{x,y,z\}$ is
used in arriving at Eq. (\ref{RDgauge}). The unified spin-dependent gauge
$\mathbf{A=A}^{B}I_{s}+\mathbf{A}^{\text{SO}}$ now is no longer a scalar, and
its components, in general, are non-commutative,
\begin{equation}
\left[  A_{x},A_{y}\right]  =2i\times\left(  \frac{mc}{e\hbar}\right)
^{2}\left(  \alpha^{2}-\beta^{2}\right)  \sigma_{z}. \label{com_tion}%
\end{equation}

Although the presence of the non-Abelian gauge field $\mathbf{A}^{\text{SO}}$,
usually referred to as the Yang-Mills field,\cite{YMfield} makes the usual
scalar gauge formalism incapable and thus complicates the problem, the
exception to the commutation Eq. (\ref{com_tion}) is clearly seen when RSO and
DSO couplings are of equal strengths $\left\vert \alpha\right\vert =\left\vert
\beta\right\vert $. In addition, for the one-dimensional system, since only
one component of the gauge $\mathbf{A}^{\text{SO}}$ will be introduced via Eq.
(\ref{RDgauge}), the non-commuatative relation Eq. (\ref{com_tion}) will not
necessarily be confronted. In Sec. \ref{Acon}, we will show that both cases
can be transformed into the free electron gas. In particular, for $\left\vert
\alpha\right\vert =\left\vert \beta\right\vert $ in Rashba-Dresselhaus [001]
model, this transformation is a gauge transformation that simplifies the
physical description of the PSH.\cite{PSHSCZ, PSHHaoger}

\subsection{Discrete case: Tight-binding model\label{Hdis}}

Conventional TB model\cite{Grosso} is based on the real crystal structure.
Assuming that each electron orbital is localized around its associated nuclei,
all spatial interactions involving more than three nuclei centers can thus be
neglected. In other words, considering only the nearest-neighbor hopping
$t_{\mathbf{m}\sigma\mathbf{m}^{\prime}\sigma^{\prime}}$, from site
$\mathbf{m}^{\prime}$ with spin $\sigma^{\prime}$ to site $\mathbf{m}$ with
spin $\sigma$, will be sufficient and is the case in the present work. The
Hamiltonian can therefore be represented in the form%
\begin{equation}
\mathcal{H}^{\text{TB}}=\sum_{\mathbf{m}\sigma}u_{\mathbf{m}\sigma
}c_{\mathbf{m}\sigma}^{\dagger}c_{\mathbf{m}\sigma}+\sum
_{\substack{\left\langle \mathbf{m},\mathbf{m}^{\prime}\right\rangle
\\\sigma\sigma^{\prime}}}t_{\mathbf{m}\sigma\mathbf{,m}^{\prime}\sigma
^{\prime}}c_{\mathbf{m}\sigma}^{\dagger}c_{\mathbf{m}^{\prime}\sigma^{\prime}%
}, \label{HTB}%
\end{equation}
with the on-site energy $u_{\mathbf{m}\sigma}$ tunable by the gate voltage,
and the fermion creation (annihilation) operators being denoted by
$c_{\mathbf{m}\sigma}$ ($c_{\mathbf{m}^{\prime}\sigma^{\prime}}^{\dagger}$),
obeying $\{c_{\mathbf{m}\sigma},c_{\mathbf{m}^{\prime}\sigma^{\prime}%
}^{\dagger}\}=1$.

On the other hand, the FD method\cite{Databook} approximates differentiations
by discretizing variables. For example, the differentiation $p_{x}\psi(x)$,
from the momentum operator $p_{x}$, is replaced with $-i\hbar(\psi_{x+a}%
-\psi_{x-a})/a\cong-i\hbar d\psi(x)/dx$. Clearly, this approximation is valid
only when the wave function $\psi\left(  x\right)  $ varies slowly over one
lattice constant $a$, or under the condition $a\ll1$. Physically, this is the
case in which only electrons near the band bottom, where $p=\hbar k$ is small,
enter our problems. Despite such a restriction, this method is still quite
general and powerful since it can deal with any single-particle Hamiltonian by
a matrix form so that the problem can be solved by numerical method. Even
though the FD method is based on different physical assumptions than the
conventional TB model, the single-particle Hamiltonian can always be expressed
in the form\cite{Databook} of Eq. (\ref{HTB}). In particular, it allows us to
make the continuos Hamiltonian Eq. (\ref{RDha}) a connection to the discrete
SO-interacting TB model.

Consider the free-electron system,
\begin{equation}
H^{\text{free}}=\frac{\mathbf{\Pi}^{2}}{2m}\text{.} \label{HfreenoV}%
\end{equation}
The TB Hamiltonian corresponding to Eq. (\ref{HfreenoV}) reads\cite{Databook},%
\begin{equation}
\mathcal{H}^{\text{free}}=\sum_{\mathbf{m}\sigma}c_{\mathbf{m}\sigma}%
^{\dagger}c_{\mathbf{n}\sigma}u_{\mathbf{m}\sigma}^{\text{free}}%
+\sum_{\substack{\left\langle \mathbf{m},\mathbf{m}^{\prime}\right\rangle
\\\sigma\sigma^{\prime}}}t_{\mathbf{m}\sigma\mathbf{,m}^{\prime}\sigma
^{\prime}}^{\text{free}}c_{\mathbf{m}\sigma}^{\dagger}c_{\mathbf{m}^{\prime
}\sigma^{\prime}}\text{.} \label{TBHfreenoV}%
\end{equation}
Each dimension contributes the quantity $2t_{0}I_{s}$ to the on-site energy,
and in a square lattice we therefore have%
\begin{equation}
u_{\mathbf{m}}^{\text{free}}=4t_{0}I_{s}\text{,} \label{freeU}%
\end{equation}
with $u_{\mathbf{m}\sigma}^{\text{free}}=\langle\sigma|u_{\mathbf{m}%
}^{\text{free}}|\sigma\rangle$, the hopping strength defined as $t_{0}%
=\hbar^{2}/2ma^{2}$, and the spin identity matrix denoted by $I_{s}$. The
magnetic gauge $\mathbf{A}^{B}$ is introduced as a phase factor in the hopping
matrix%
\begin{equation}
t_{\mathbf{m,m}^{\prime}}^{\text{free}}=-t_{0}\exp\left[  \frac{ie}{c\hbar
}\mathbf{A}^{B}\cdot\left(  \mathbf{m}-\mathbf{m}^{\prime}\right)  \right]
I_{s}\text{,} \label{freet}%
\end{equation}
with $t_{\mathbf{m}\sigma\mathbf{,m}^{\prime}\sigma^{\prime}}^{\text{free}%
}=\langle\sigma|t_{\mathbf{m,m}^{\prime}}^{\text{free}}|\sigma^{\prime}%
\rangle$. We note that the phase acquired by an electron hopping from one site
to another over a distance $a$ is proportional to $a\mathbf{A}^{B}$. If we
analogize the $\mathbf{A}^{\text{SO}}$ as $\mathbf{A}^{B}$ in Eq.
(\ref{freet}), and expand the \textit{SU}(2) phase as%
\begin{equation}
e^{\frac{ie}{c\hbar}A_{x}^{\text{SO}}a}e^{\frac{ie}{c\hbar}A_{y}^{\text{SO}}%
a}\approx e^{i\frac{ie}{c\hbar}\left(  A_{x}^{\text{SO}}+A_{y}^{\text{SO}%
}\right)  a}+O\left\{  \left(  \frac{ea}{c\hbar}\right)  ^{2}\left[
A_{x}^{SO},A_{y}^{SO}\right]  \right\}  \text{,} \label{AxAyapp}%
\end{equation}
with $[A_{x}^{\text{SO}},A_{y}^{\text{SO}}]\equiv A_{x}^{\text{SO}}%
A_{y}^{\text{SO}}-A_{y}^{\text{SO}}A_{x}^{\text{SO}}$, we find that in the
limit
\begin{equation}
\frac{e}{c\hbar}\mathbf{A}^{\text{SO}}\cdot\left(  \mathbf{m}-\mathbf{m}%
^{\prime}\right)  \ll1\text{,} \label{FDcond}%
\end{equation}
the SO gauge $\mathbf{A}^{\text{SO}}$ plays a similar role as $\mathbf{A}^{B}$
due to their equivalent algebra $\exp[(iea/c\hbar)(A_{x}^{B\text{(SO)}}%
)]\exp[(iea/c\hbar)(A_{x}^{B\text{(SO)}})]=\exp[(iea/c\hbar)(A_{x}%
^{B\text{(SO)}}+A_{y}^{B\text{(SO)}})]$. Accordingly, in the FD approximation
with lattice spacing small enough to satisfy Eq. (\ref{FDcond}), the
replacements of
\begin{equation}
4t_{0}\rightarrow4t_{0}-V\text{,} \label{trep}%
\end{equation}
in Eq. (\ref{freeU}) and%
\begin{equation}
\mathbf{A}^{B}\rightarrow\mathbf{A}^{B}+\mathbf{A}^{\text{SO}}\text{,}
\label{Arep}%
\end{equation}
in Eq. (\ref{freet}) yield the expression, in the TB form of our SO
Hamiltonian Eq. (\ref{RDgauge}),%
\begin{equation}
\mathcal{H}=\sum_{\mathbf{m}\sigma}u_{\mathbf{m}\sigma}c_{\mathbf{m}\sigma
}^{\dagger}c_{\mathbf{n}\sigma}+\sum_{\substack{\left\langle \mathbf{m}%
,\mathbf{m}^{\prime}\right\rangle \\\sigma\sigma^{\prime}}}t_{\mathbf{m}%
\sigma\mathbf{,m}^{\prime}\sigma^{\prime}}c_{\mathbf{m}\sigma}^{\dagger
}c_{\mathbf{m}^{\prime}\sigma^{\prime}}\text{,} \label{SOTB}%
\end{equation}
with now $u_{\mathbf{m}}=(4t_{0}-V)I_{s}$ and $t_{\mathbf{m,m}^{\prime}%
}=-t_{0}\exp\left[  (ie/c\hbar)\mathbf{A}\cdot\left(  \mathbf{m}%
-\mathbf{m}^{\prime}\right)  \right]  $ taking into account both the
background potential $V$ and SO gauge $\mathbf{A}^{\text{SO}}$.

To make a comparison with the TB Hamiltonian of Ref. \onlinecite{Nicolic}
where a special case of the square lattice without any applied magnetic field
is considered, we let $\mathbf{A}^{B}=0$ and expand the hopping matrix up to
the second order of the exponent $(e/c\hbar)\mathbf{A}^{\text{SO}}\cdot\left(
\mathbf{m}-\mathbf{m}^{\prime}\right)  $,%
\begin{align}
&  t_{\mathbf{m,m}^{\prime}}\nonumber\\
&  \cong-t_{0}\left\{  1+\frac{ie}{c\hbar}\mathbf{A}^{\text{SO}}\cdot\left(
\mathbf{m}-\mathbf{m}^{\prime}\right)  -\left[  \frac{e}{c\hbar}%
\mathbf{A}^{\text{SO}}\cdot\left(  \mathbf{m}-\mathbf{m}^{\prime}\right)
\right]  ^{2}\right\} \nonumber\\
&  =\left\{
\begin{array}
[c]{cc}%
-t_{0}\left[  1+i\dfrac{t^{R}}{t_{0}}\sigma_{y}-i\dfrac{t^{D}}{t_{0}}%
\sigma_{x}\right]  +\dfrac{V}{2} & \text{for }\mathbf{m}=\mathbf{m}^{\prime
}+a\mathbf{e}_{x}\\
-t_{0}\left[  1-i\dfrac{t^{R}}{t_{0}}\sigma_{x}+i\dfrac{t^{D}}{t_{0}}%
\sigma_{y}\right]  +\dfrac{V}{2} & \text{for }\mathbf{m}=\mathbf{m}^{\prime
}+a\mathbf{e}_{y}%
\end{array}
\right.  \text{,} \label{expandt}%
\end{align}
where the RSO and DSO hoppings are respectively defined by $t^{R}=\alpha/2a$
and $t^{D}=\beta/2a$. Moreover, in such limit Eq. (\ref{FDcond}), we can
approximate the operation $-V\psi_{\mathbf{m}}$ as $-V\left(  \psi
_{\mathbf{m+}a\mathbf{e}_{i}}+\psi_{\mathbf{m-}a\mathbf{e}_{i}}\right)  /2$
with $i\in\left\{  x,y\right\}  $ due to the slow variation of the wave
function $\psi_{\mathbf{m}}\cong\left(  \psi_{\mathbf{m+}a\mathbf{e}_{i}}%
+\psi_{\mathbf{m-}a\mathbf{e}_{i}}\right)  /2$. This approximation converts
the on-site background potential $-V$ into the hopping $-V/2$ and thus cancels
out the last term $V/2$ in Eq. (\ref{expandt}). Finally, in Eq. (\ref{SOTB}),
we have $u_{\mathbf{m}}=4t_{0}I_{s}$, $t_{\mathbf{m,m}^{\prime}}=$
$-t_{0}\left[  1+i(t^{R}/t_{0})\sigma_{y}-i(t^{D}/t_{0})\sigma_{x}\right]  $
\ for $\mathbf{m}=\mathbf{m}^{\prime}+a\mathbf{e}_{x}$ and $t_{\mathbf{m,m}%
^{\prime}}=$ $-t_{0}\left[  1-i(t^{R}/t_{0})\sigma_{x}+i(t^{D}/t_{0}%
)\sigma_{y}\right]  $ for $\mathbf{m}=\mathbf{m}^{\prime}+a\mathbf{e}_{y}$. By
further letting $t^{D}=0$, Eq. (\ref{SOTB}) reduces to the form adopted in
Ref. \onlinecite {Nicolic}. This suggests that, in the FD approximation under
condition Eq. (\ref{FDcond}), it is reasonable to treat the SO gauge as a
magnetic one.

For the 2DEG made of InGaAs/InAlAs heterostructure grown along [001]
direction, typical parameters,\cite{Nitta} the effective electron mass
$m\approx0.05m_{e}$ ($m_{e}$ is the electron mass), the SO coupling strength
$\approx0.3%
\operatorname{eV}%
\operatorname{\text{\AA}}%
$ and the lattice constant $a\approx3%
\operatorname{nm}%
$, yield the considerably small value $(e/c\hbar)\mathbf{A}^{\text{SO}}%
\cdot\left(  \mathbf{m}-\mathbf{m}^{\prime}\right)  \approx1.5\times10^{-3}$.
Therefore, Eq. (\ref{FDcond}) is indeed realizable. Moreover, if only Eq.
(\ref{FDcond}) is satisfied, the \textit{SU}(2) gauge can be approximated as
the \textit{U}(1) gauge. Two generalities of the TB Hamiltonian Eq.
(\ref{SOTB}) are then worth mentioning. First, the sites $\mathbf{m}$ are not
necessarily arranged on a square lattice, i.e., one can apply Eq. (\ref{SOTB})
to the hopping along an arbitrary direction. Second, for spatial-dependent
(nonuniform) SO interactions $\alpha=\alpha\left(  \mathbf{r}\right)  $ and
$\beta=\beta\left(  \mathbf{r}\right)  $, the hopping exponent $\mathbf{A}%
^{\text{SO}}\cdot\left(  \mathbf{m}-\mathbf{m}^{\prime}\right)  $ can simply
be replaced with $\int_{\mathbf{m}^{\prime}}^{\mathbf{m}}\mathbf{A}%
^{\text{SO}}\left(  \mathbf{r}^{\prime}\right)  \cdot d\mathbf{r}^{\prime}$
and the formalism will be the same.

\section{Applications\label{SA}}

We present here two applications of the continuous case, in Sec. \ref{Acon},
based on Eq. (\ref{RDgauge}) and the discrete case, in Sec. \ref{Adis}, based
on Eq. (\ref{SOTB}). The former studies the PSH from the gauge transformation,
while the latter extends the previous work of Ref. \onlinecite{Non-Abelian} on
the quantum square ring.

\subsection{Continuos case: Gauge transformation and persistent spin
helix\label{Acon}}

Consider the local transformation operator $U\left(  \mathbf{r}\right)
=\exp[(ie/\hbar c)\int_{c}\mathbf{A}^{\text{SO}}\left(  \mathbf{r}^{\prime
}\right)  \cdot d\mathbf{r}^{\prime}]$. In general, $U\left(  \mathbf{r}%
\right)  $ depends on the actual integration path $c$. Nevertheless, in the
case of uniform Rashba and Dresselhaus case, since $\mathbf{A}^{\text{SO}}$ is
independent of position $\mathbf{r=(}x,y)$, we have $\nabla\times
\mathbf{A}^{\text{SO}}=0$, so that $\int_{c}\mathbf{A}^{\text{SO}}%
d\mathbf{r}^{\prime}=\mathbf{A}^{\text{SO}}\cdot\mathbf{r}$ becomes
path-independent. This yields the expression%
\begin{equation}
U\left(  \mathbf{r}\right)  =\exp\left[  \frac{ie}{\hbar c}\left(
\mathbf{A}^{\text{SO}}\cdot\mathbf{r}\right)  \right]  \text{,} \label{Udef}%
\end{equation}
with the unitary property $U\left(  \mathbf{r}\right)  U^{\dagger}\left(
\mathbf{r}\right)  =I_{s}$ ensured by the Hermitian $\mathbf{A}^{\text{SO}%
\dagger}=\mathbf{A}^{\text{SO}}$ from definition Eq. (\ref{SOgauge}). We
notice that Eq. (\ref{RDgauge}) differs from the free electron gas (with a
background potential $V$),%
\begin{equation}
h=\frac{\mathbf{\Pi}^{2}}{2m}-V\text{,} \label{hfree}%
\end{equation}
only by a gauge $(e/c)\mathbf{A}^{\text{SO}}$. This suggests to consider the
transformation,
\begin{align}
&  U\left(  \mathbf{r}\right)  \mathbf{\Pi}U^{\dagger}\left(  \mathbf{r}%
\right) \nonumber\\
&  =\mathbf{\Pi+}\frac{ie}{\hbar c}\left[  \mathbf{A}^{\text{SO}}%
\cdot\mathbf{r,\Pi}\right] \nonumber\\
&  +\frac{1}{2}\left(  \frac{ie}{\hbar c}\right)  ^{2}\left[  \mathbf{A}%
^{\text{SO}}\cdot\mathbf{r,}\left[  \mathbf{A}^{\text{SO}}\cdot\mathbf{r,\Pi
}\right]  \right]  +\cdots, \label{Transform}%
\end{align}
with $[\mathbf{A}^{\text{SO}}\cdot\mathbf{r,\Pi]}=i\hbar\mathbf{A}^{\text{SO}%
}$. Unsatisfactorily, due to the non-commutability $[A_{x}^{\text{SO}}%
,A_{y}^{\text{SO}}]\neq0$, the higher order terms in Eq. (\ref{Transform}), in
general, do not vanish, leading to $U\left(  \mathbf{r}\right)  hU^{\dagger
}\left(  \mathbf{r}\right)  \neq H$.

Exceptionally, in the equal-strength case $\left\vert \alpha\right\vert
=\left\vert \beta\right\vert $, $A_{x}^{\text{SO}}$ and $A_{y}^{\text{SO}}$
follow the scalar algebra $A_{x}^{\text{SO}}A_{y}^{\text{SO}}=A_{y}%
^{\text{SO}}A_{x}^{\text{SO}}$, and hence we obtain $U\left(  \mathbf{r}%
\right)  \mathbf{\Pi}U^{\dagger}\left(  \mathbf{r}\right)  =\mathbf{\Pi
-}e/c\mathbf{A}^{\text{SO}}$, or the gauge transformation,%
\begin{equation}
U^{\dagger}\left(  \mathbf{r}\right)  HU\left(  \mathbf{r}\right)  =h.
\label{GaugeTrans}%
\end{equation}
As a result, the free electron gas $h$ in Eq. (\ref{hfree}) and the
SO-interacting electron gas $H$ in Eq. (\ref{RDgauge}) share the same
eigenenergies $E_{\mathbf{k}}$. Their corresponding eigenfunctions, denoted by
$\psi_{E_{\mathbf{k}}}(\mathbf{r)}\chi_{s}^{\text{free}}$ and $\Psi
_{E_{\mathbf{k}}}(\mathbf{r)}\chi_{s}^{\text{SO}}$, respectively, differ from
each other only by a phase factor from the $2\times2$ matrix $U(\mathbf{r)}$,
namely, $\Psi_{E_{\mathbf{k}}}(\mathbf{r)}\chi_{s}^{\text{SO}}=$
$U(\mathbf{r)}\psi_{E_{\mathbf{k}}}(\mathbf{r)}\chi_{s}^{\text{free}}$.
Moreover, any wave function is constructed by a superposition of the
eigenfunctions, so that for any given wave function $\psi(\mathbf{r)}\chi
_{s}^{\text{free}}$ in $h$, the corresponding wave function in $H$ is
$U(\mathbf{r)}\psi(\mathbf{r)}\chi_{s}^{\text{free}}$. Below, we show that the
physical description of the PSH in the SO interacting system can be easily
understood by this correspondence.

Consider first an injected electron in system $h$, described by $\psi
_{\text{inj}}(\mathbf{r)}\chi_{\text{inj}}=[\sum_{\mathbf{k}}C_{\mathbf{k}%
}\psi_{E_{\mathbf{k}}}(\mathbf{r)]}\chi_{\text{inj}}$, with the initial spin
state $\chi_{\text{inj}}$ and the weight factor $C_{\mathbf{k}}$. Clearly,
without any spin-dependent mechanisms, this electron shall retain its spin
state $\chi_{\text{inj}}$ as it traverses the sample. Now, import
$\mathbf{A}^{\text{SO}}$. This corresponds to turning on $U(\mathbf{r)}$ so
that the electron wave function, in the SO-interacting system $H$, undergoes a
gauge transformation of $U(\mathbf{r)}$,
\begin{equation}
U\left(  \mathbf{r}\right)  \psi_{\text{inj}}\left(  \mathbf{r}\right)
\chi_{\text{inj}}=\sum_{\mathbf{k}}C_{\mathbf{k}}\psi_{E_{\mathbf{k}}}\left(
\mathbf{r}\right)  U\left(  \mathbf{r}\right)  \chi_{\text{inj}}\text{.}
\label{SOinj}%
\end{equation}
Accordingly, the spin polarization of the electron varies spatially according
to $U(\mathbf{r)}\chi_{\text{inj}}$. Using Eqs. (\ref{Udef}) and
(\ref{SOgauge}), with $\alpha=\beta$, we find that%
\begin{align}
\left.  U\left(  \mathbf{r}\right)  \right\vert _{\alpha=\beta}  &
=\exp\left[  -\alpha\frac{i2m}{\hbar^{2}}\sigma_{\left(  1,-1\right)
}r_{\left(  1,1\right)  }\right] \nonumber\\
&  =\exp\left[  -i\frac{\hbar\sigma_{\left(  1,-1\right)  }/2}{\hbar}%
\theta_{\text{PSH}}^{+}\right]  , \label{Ua=b}%
\end{align}
with $\sigma_{\left(  1,-1\right)  }\equiv(\sigma_{x},\sigma_{y}%
)\cdot(1,-1)/\sqrt{2}$ is actually the spin rotation operator, with rotation
axis along $(1,-1)$ and the precession angle $\theta_{\text{PSH}}^{+}%
\equiv(4\alpha m/\hbar^{2})r_{\left(  1,1\right)  }$ depending on the distance
$r_{\left(  1,1\right)  }\equiv\mathbf{r}\cdot(1,1)/\sqrt{2}$ along $(1,1)$
[cf. Fig 2(b) in Ref. \onlinecite{PSHHaoger}].

Similarly, for $\alpha=-\beta$, we have
\begin{equation}
\left.  U\left(  \mathbf{r}\right)  \right\vert _{\alpha=-\beta}=\exp\left[
-i\frac{\hbar\sigma_{\left(  1,1\right)  }/2}{\hbar}\theta_{\text{PSH}}%
^{-}\right]  , \label{Ua=-b}%
\end{equation}
corresponding to a rotation axis along $(1,1)$ and a precession angle
$\theta_{\text{PSH}}^{-}=(4\alpha m/\hbar^{2})r_{\left(  -1,1\right)  }$ with
$r_{\left(  -1,1\right)  }\equiv\mathbf{r}\cdot(-1,1)/\sqrt{2}$. This is
precisely the PSH with precession length,
\begin{equation}
L_{P}=\frac{\hbar^{2}\pi}{2m\alpha}\text{,} \label{Lp}%
\end{equation}
for spin to rotate $\theta_{\text{PSH}}^{+}$ or $\theta_{\text{PSH}}^{-}=2\pi
$. Obviously, the PSH is robust against any spin-independent mechanisms for
$U^{\dagger}\left(  \mathbf{r}\right)  V_{p}\left(  \mathbf{r}\right)
U\left(  \mathbf{r}\right)  =V_{p}\left(  \mathbf{r}\right)  $ which holds for
any potential of the form $V_{p}\left(  \mathbf{r}\right)  \propto I_{s}$,
including the finite-size confinement due to spin-independent boundaries. In
particular, the presence of $\mathbf{A}^{B}$, which determines the actual form
of $\psi_{E_{\mathbf{k}}}\left(  \mathbf{r}\right)  $, clearly does not affect
the PSH. This is reasonable since $\mathbf{A}^{B}$ contributes only a
spin-independent phase to the electron wavefunction due to the absence of the
Zeeman term, and therefore does not vary the spin polarization. Inclusion of
the Zeeman term requires further generalization, which is beyond the scope of
the present discussion.

A similar calculation can also apply to the DSO [110] linear model,%
\begin{equation}
H_{\left[  110\right]  }=\frac{p_{x}^{2}+p_{y}^{2}}{2m}-\frac{2\beta}{\hbar
}p_{x}\sigma_{z}\text{,} \label{110H}%
\end{equation}
with the gauge $\mathbf{A}_{\left[  110\right]  }^{\text{SO}}=(A_{x\left[
110\right]  },A_{y\left[  110\right]  },0)\equiv(2\beta\sigma_{z}%
,0,0)mc/e\hbar$. Due to $[A_{x\left[  110\right]  },A_{y\left[  110\right]
}]=0$, we also have here the PSH described by
\begin{equation}
U\left(  \mathbf{r}\right)  _{\left[  110\right]  }=\exp\left(  -i\frac
{\hbar\sigma_{z}/2}{\hbar}\theta_{\text{PSH}}\right)  , \label{110U}%
\end{equation}
with $\theta_{\text{PSH}}=-4m\beta x/\hbar^{2}$ [cf. Fig. 2(d) in Ref. \onlinecite{PSHHaoger}].

We further point out here that the one-dimensional SO-interacting system (with
any values of $\alpha$ and $\beta$), also possesses the applicability of this
transformation. For example, consider a one-dimensional (along $\mathbf{e}%
_{n}$ direction) SO-interacting conductor. Since the degrees of freedom of
orthogonal directions are frozen, only $\Pi_{n}\equiv\mathbf{\Pi\cdot e}_{n}$
and $A_{n}^{\text{SO}}\equiv$ $\mathbf{A}^{\text{SO}}\cdot\mathbf{e}_{n}$ are
relevant to our problem and appear in Eq. (\ref{RDgauge}). The system can now
be described by $H^{\text{1D}}=[\Pi_{n}\mathbf{-(}e/c)A_{n}^{\text{SO}}%
]^{2}/(2m)-V^{\text{1D}}$, with $V^{\text{1D}}$ being a constant potential and
the non-commutability Eq. (\ref{com_tion}) is not encountered. Following
procedure similar to those above, we arrive again at the transformation
$U^{\text{1D}\dagger}(r_{n})H^{\text{1D}}U^{\text{1D}}(r_{n})=h^{\text{1D}%
}=\Pi_{n}^{2}/2m-V^{\text{1D}}$, with $U^{\text{1D}}\left(  r_{n}\right)
=\exp[(ie/\hbar c)A_{n}^{\text{SO}}\cdot r_{n}]$ with $r_{n}\equiv
\mathbf{r\cdot e}_{n}$.

\subsection{Discrete case: Four states in the quantum square ring\label{Adis}}

\begin{figure}
[tb]
\begin{center}
\includegraphics[
height=1.3465in,
width=3.3641in
]%
{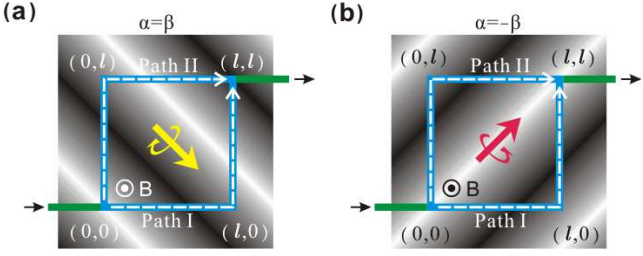}%
\caption{(Color online) The square ring cornering at $(0,0),(l,0),(l,l)$ and
$(l,l)$, in contact with two leads. Magnetic field $\mathbf{B}=B\mathbf{e}%
_{z}$ is locally applied at the center of the ring. Electrons are injected
from the left lead, interfered via path I and path II, and then transmitted to
the right lead. The background shading represents the helix precession angle.
From white to black, the electron precesses by $\pi$. The rotation axes are
specifed by the arrows at the centers of the rings for (a) $\alpha=\beta$ and
(b) $\alpha=-\beta$.}%
\label{fig1}%
\end{center}
\end{figure}

Using the TB Hamiltonian of Eq. (\ref{SOTB}), we investigate, in this section,
the square ring interferometer\cite{Non-Abelian} (see Fig. \ref{fig1}) under
the interactions of the magnetic field, DSO and RSO couplings.%
The transmission coefficients $T_{\sigma^{\prime}\sigma}$, with incoming
(outgoing) spin denoted by $\sigma$($\sigma^{\prime}$), depends on the
interplay among the phases due to the RSO, DSO interactions and the external
field. Based on different values of the magnetic field $B$, the Rashba
coefficient $\alpha$, and Dresselhaus coefficient $\beta$, four states in this
setup: "perfect" insulating, spin-filtering,\cite{Spin Filter1, Spin Filter2,
Non-Abelian} spin-flipping, and spin-keeping states, will be identified. When
electrons are injected into the ring, the spin-insulating ring blocks both up
and down spin channels, the spin-filtering ring blocks only one of the spin
channels, the spin-flipping ring flips the spins of the injected electrons,
while the spin-keeping ring keeps the injected spin configuration unaltered.
It is worth mentioning that these four states are valid for any range of the
injected energy $E$ as the term "perfect" stands for.

Consider a square ring (see Fig. \ref{fig1}) described by $\mathcal{H}$ in Eq.
(\ref{SOTB}), with width $l$ and cornering at $\mathbf{m}=(0,0),$ $(l,0),$
$(l,l)$, and $(0,l)$. The magnetic field $\mathbf{B}=B\mathbf{e}_{z}$
penetrating this ring is applied only locally so that the Zeeman term does not
emerge. Two ideal leads contact the ring at the positions $\left(  0,0\right)
$ and $\left(  l,l\right)  $. For brevity, and without loss of generality, we
choose the on-site energy in both leads and the ring to be zero. The
self-energy\cite{Databook} $\Sigma(E)=(E-i\sqrt{4t_{0}^{2}-E^{2}})(c_{\left(
0,0\right)  }^{\dagger}c_{\left(  0,0\right)  }+c_{\left(  l,l\right)
}^{\dagger}c_{\left(  l,l\right)  })/2$ is generated due to the presence of
the leads. The transmission coefficients are computed by%
\begin{equation}
T_{\sigma^{\prime}\sigma}=\left\langle \left(  l,l\right)  ;\sigma^{\prime
}\right\vert \frac{\sqrt{4t_{0}^{2}-E^{2}}}{E-\mathcal{H-}\Sigma\left(
E\right)  }\left\vert \left(  0,0\right)  ;\sigma\right\rangle \text{,}%
\label{Texpression}%
\end{equation}
with the incoming (from left lead) spin $\sigma$ and outgoing (to the right
lead) spin $\sigma^{\prime}$ states denoted as $\left\vert \left(  0,0\right)
;\sigma\right\rangle $ and $\left\vert \left(  l,l\right)  ;\sigma^{\prime
}\right\rangle $, respectively. The interference between the bottom-right path
I $\equiv(0,0)\rightarrow(l,0)\rightarrow(l,l)$ and the left-top path II
$\equiv(0,0)\rightarrow(0,l)\rightarrow(l,l)$ is determined by the phase,
circling the ring, of the form $U_{\text{phase}}=U_{\text{II}}^{\dagger
}U_{\text{I}}=[U_{\left(  l,l\right)  \leftarrow\left(  0,l\right)
}U_{\left(  0,l\right)  \leftarrow\left(  0,0\right)  }]^{\dagger}U_{\left(
l,l\right)  \leftarrow\left(  l,0\right)  }U_{\left(  l,0\right)
\leftarrow\left(  0,0\right)  }=U_{\left(  0,0\right)  \leftarrow\left(
0,l\right)  }U_{\left(  0,l\right)  \leftarrow\left(  l,l\right)  }U_{\left(
l,l\right)  \leftarrow\left(  l,0\right)  }U_{\left(  l,0\right)
\leftarrow\left(  0,0\right)  }$ with $U_{\mathbf{m}^{\prime}\leftarrow
\mathbf{m}}\equiv\exp[(ie/c\hbar\mathbf{)A}\cdot(\mathbf{m}-\mathbf{m}%
^{\prime})]$ and $\mathbf{A=A}^{B}I_{s}+\mathbf{A}^{\text{SO}}$ accounting for
the magnetic gauge $\mathbf{A}^{B}=(-By,Bx,0)/2$ and the SO gauge Eq.
(\ref{SOgauge}). Explicitly, in the square ring, we obtain the 2 by 2 matrix
\begin{align}
U_{\text{phase}} &  =e^{i\phi^{B}}e^{i\frac{ml}{\hbar^{2}}\left(  \alpha
\sigma_{x}-\beta\sigma_{y}\right)  }e^{-i\frac{ml}{\hbar^{2}}\left(
\alpha\sigma_{y}-\beta\sigma_{x}\right)  }\nonumber\\
&  \times e^{-i\frac{ml}{\hbar^{2}}\left(  \alpha\sigma_{x}-\beta\sigma
_{y}\right)  }e^{i\frac{ml}{\hbar^{2}}\left(  \alpha\sigma_{y}-\beta\sigma
_{x}\right)  }\text{,}\label{Uphase}%
\end{align}
corresponding to the magnetic flux $\phi^{B}=(e/c\hbar)Bl^{2}$. Denote the
general spin-up and -down along the direction with the polar angle $\theta
\in\lbrack0,\pi]$ and azimuthal angle $\phi\in\lbrack0,2\pi]$, as
$\uparrow_{\left[  \theta,\phi\right]  }$ and $\downarrow_{\left[  \theta
,\phi\right]  }$ respectively, i.e.,
\begin{equation}
\uparrow_{\left[  \theta,\phi\right]  }=\binom{e^{-i\phi/2}\cos\dfrac{\theta
}{2}}{e^{i\phi/2}\sin\dfrac{\theta}{2}}\text{.}\label{gspindef}%
\end{equation}
Strictly speaking, in the arbitrary $\uparrow_{\left[  \theta,\phi\right]  }$
and $\downarrow_{\left[  \theta,\phi\right]  }$ axes, $U_{\text{phase}}$ does
not give us the meaning of "phase", since it is in general a non-diagonal
matrix, while in the diagonal axes with $DU_{\text{phase}}D^{\dagger}=\Lambda
$, defining the tilted up $\left\vert \tilde{\uparrow}\right\rangle $ (down
$\left\vert \tilde{\downarrow}\right\rangle $) spin as $\left\vert
\tilde{\uparrow}\right\rangle =D|\uparrow_{\left[  0,0\right]  }\rangle$
($\left\vert \tilde{\downarrow}\right\rangle =D|\downarrow_{\left[
0,0\right]  }\rangle$), the circled phase $\phi_{\tilde{\uparrow}}$ acquired
by $\left\vert \tilde{\uparrow}\right\rangle $ ($\left\vert \tilde{\downarrow
}\right\rangle $) electron can be read off via the eigenvalues of
$U_{\text{phase}}$,
\begin{equation}
\Lambda=\left(
\begin{array}
[c]{cc}%
e^{i\phi_{\tilde{\uparrow}}} & 0\\
0 & e^{i\phi_{\tilde{\downarrow}}}%
\end{array}
\right)  \text{,}\label{lambda}%
\end{equation}
with $\phi_{\tilde{\uparrow}\left(  \tilde{\downarrow}\right)  }\equiv\phi
^{B}+\phi_{_{\tilde{\uparrow}\left(  \tilde{\downarrow}\right)  }}^{\text{SO}%
}$ contributed from both magnetic $\phi^{B}$ and SO $\phi_{_{\tilde{\uparrow
}\left(  \tilde{\downarrow}\right)  }}^{\text{SO}}$ phases. Consider a special
case with vanishing $\beta$, $\phi^{B}=\pi/2$, and $\phi_{\tilde{\uparrow
}\left(  \tilde{\downarrow}\right)  }^{\text{SO}}=$ $+\left(  -\right)  \pi
/2$. This corresponds to the Rashba strength\cite{Non-Abelian},%
\begin{equation}
\alpha^{\ast}=\frac{\hbar^{2}}{ml}\sin^{-1}\left(  2^{-1/4}\right)
\text{.}\label{alphastar}%
\end{equation}
There is, therefore, destructive interference $e^{i\phi_{\tilde{\uparrow}}%
}=e^{i\pi}=-1$ for $\left\vert \tilde{\uparrow}\right\rangle $ electrons, and
constructive interference $e^{i\phi_{\tilde{\downarrow}}}=e^{i0}=1$ for
$\left\vert \tilde{\downarrow}\right\rangle $ electrons. The spin-filtering
ring (filtering out $\left\vert \tilde{\uparrow}\right\rangle $) is thus
achieved in this case. We now numerically analyze the SO phase $\phi
_{_{\tilde{\uparrow}\left(  \tilde{\downarrow}\right)  }}^{\text{SO}}$ and the
transmission coefficients $T_{\sigma^{\prime}\sigma}$ in Eq.
(\ref{Texpression}).%
\begin{figure}
[tb]
\begin{center}
\includegraphics[
height=2.2632in,
width=3.3633in
]%
{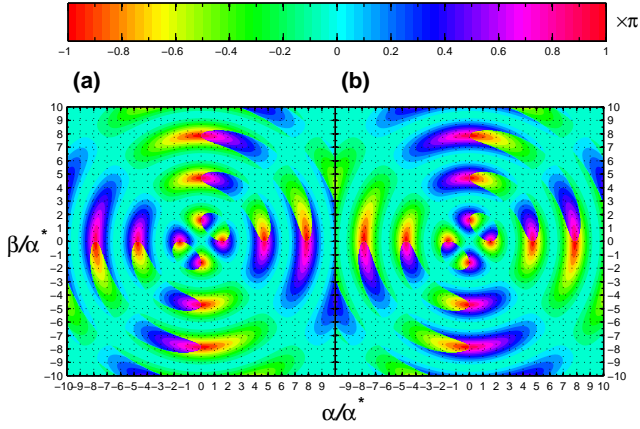}%
\caption{(Color online) Spin-orbit phases in (a) $\phi_{_{\tilde{\uparrow}}%
}^{\text{SO}}$ and (b) $\phi_{_{\tilde{\downarrow}}}^{\text{SO}}$ as functions
of $\alpha/\alpha^{\ast}$ and $\beta/\alpha^{\ast}$ for the square ring
patterned in the Rashba-Dresselhaus [001] 2DEG.}%
\label{fig2}%
\end{center}
\end{figure}

In Figs. \ref{fig2}(a) and \ref{fig2}(b), we plot the tilted-spin-up phase
$\phi_{_{\tilde{\uparrow}}}^{\text{SO}}$ and down phase $\phi_{\tilde
{\downarrow}}^{\text{SO}}$ as functions of the Rashba $\alpha/\alpha^{\ast}$
and Dresselhaus $\beta/\alpha^{\ast}$ interaction strengths, with
$\alpha^{\ast}$ defined in Eq. (\ref{alphastar}). Since $U_{\text{phase}%
}\left(  \phi_{B}\rightarrow0\right)  $ in Eq. (\ref{Uphase}) is a rotation in
the \textit{SU}(2) group, we have $\det[U_{\text{phase}}(\phi_{B}%
\rightarrow0)]=\det\Lambda=1$, yielding $\phi_{_{\tilde{\uparrow}}}%
^{\text{SO}}=-\phi_{\tilde{\downarrow}}^{\text{SO}}$ which is clearly seen in
Fig. \ref{fig2}. Notice that the SO phases are symmetric\cite{Footnote}
(unchanged) under the replacement $(\alpha,\beta)\rightarrow(-\alpha,-\beta)$
but anti-symmetric (different by a sign) under $(\alpha,\beta)\rightarrow
(\beta,\alpha)$. For $\left\vert \alpha\right\vert =\left\vert \beta
\right\vert $, $U_{\text{phase}}(\phi_{B}\rightarrow0,\left\vert
\alpha\right\vert \rightarrow\left\vert \beta\right\vert )=I_{s}$ gives zero
$\phi_{\tilde{\uparrow}\left(  \tilde{\downarrow}\right)  }^{\text{SO}}$.
Moreover, in the vicinity of the circles $R_{s}=\sqrt{(\alpha^{2}+\beta
^{2})/\alpha^{\ast2}}=s\pi,$ $s=0,1,2\cdots$, we have small SO phases
$\phi_{\tilde{\uparrow}\left(  \tilde{\downarrow}\right)  }^{\text{SO}}%
\approx0$. Between these circles are the bands where the SO$\ $phases
$\phi_{_{\tilde{\uparrow}\left(  \tilde{\downarrow}\right)  }}^{\text{SO}}$
oscillate at $R_{s+1/2}$ with a full period $2\pi$ within each quadrant. The
fine structure inside the bands clearly shows discontinuities of
$\phi_{_{\tilde{\uparrow}\left(  \tilde{\downarrow}\right)  }}^{\text{SO}}$
[see, for example, the boundaries between $\phi_{_{\tilde{\uparrow}\left(
\tilde{\downarrow}\right)  }}^{\text{SO}}\approx0.4\pi$ (blue/dark) and
$\phi_{_{\tilde{\uparrow}\left(  \tilde{\downarrow}\right)  }}^{\text{SO}%
}\approx-0.4\pi$ (green/light)]. In particular, the parameter proposed for a
spin filter,\cite{Non-Abelian} $(\alpha,\beta)/\alpha^{\ast}=\pm1\times(1,0)$,
corresponding to $\phi_{\tilde{\uparrow}\left(  \tilde{\downarrow}\right)
}^{\text{SO}}=$ $+(-)\pi/2$, is quite unstable. Interestingly, the presence of
$\beta$ does help one to move $\phi_{\tilde{\uparrow}\left(  \tilde
{\downarrow}\right)  }^{\text{SO}}$ away from these discontinuities and makes
spin filter therefore experimentally more accessible, for example,
$(\alpha,\beta)/\alpha^{\ast}\cong\pm1\times(4.55,-1.8)$ in the second band
($\approx R_{3/2}$).

\begin{figure}
[ptb]
\begin{center}
\includegraphics[
height=2.7086in,
width=3.3589in
]%
{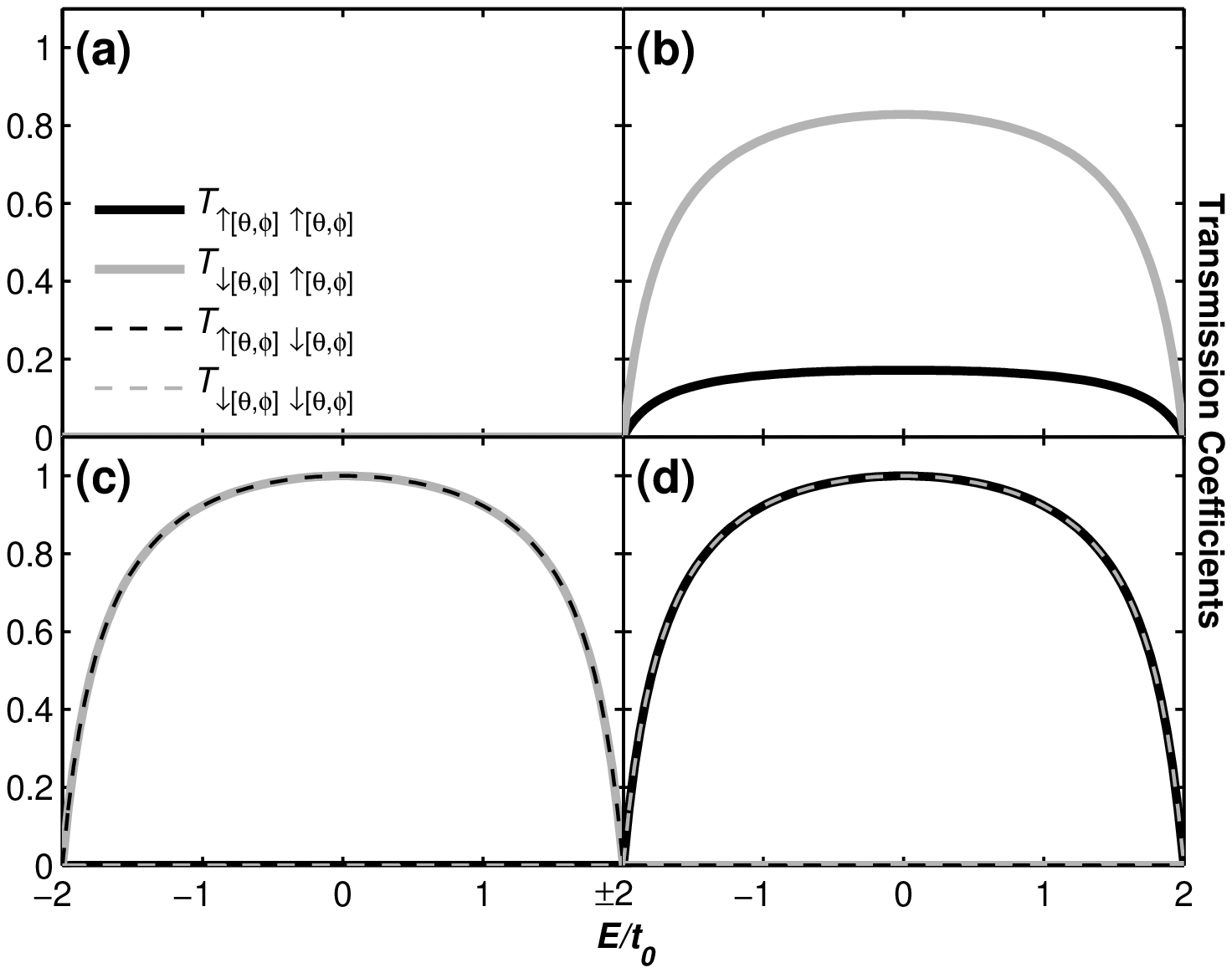}%
\caption{Transmission coefficients $T_{\uparrow_{\left[  \theta,\phi\right]
}\uparrow_{\left[  \theta,\phi\right]  }}$ (black solid lines), $T_{\downarrow
_{\left[  \theta,\phi\right]  }\uparrow_{\left[  \theta,\phi\right]  }}$(gray
solid lines), $T_{\uparrow_{\left[  \theta,\phi\right]  }\downarrow_{\left[
\theta,\phi\right]  }}$ (black dashed lines), $T_{\downarrow_{\left[
\theta,\phi\right]  }\downarrow_{\left[  \theta,\phi\right]  }}$ (gray dashed
lines) as functions of energy $E$. In the insulating state (a) and the
spin-keeping state (d), the up spin $\uparrow_{\left[  \theta,\phi\right]  }$
is defined by Eq. (\ref{gspindef}), for any $\theta\in\left[  0,\pi\right]  $
and $\phi\in\left[  0,2\pi\right]  $. In the spin-filtering state (b), up-spin
is defined in the tilted axes mentioned in Sec. \ref{Adis}, i.e.,
$\uparrow_{\left[  \theta,\phi\right]  }\equiv\tilde{\uparrow}$, whereas in
the spin-flipping state (c), it is defined in any directions lying in the
$x=y$ plane, i.e. $\uparrow_{\left[  \theta,\phi\right]  }\equiv
\uparrow_{\left[  \theta,\pi/4\right]  }$. }%
\label{fig3}%
\end{center}
\end{figure}

Figure \ref{fig3} plots the transmission coefficients $T_{\sigma^{\prime
}\sigma}$, with $\sigma$, $\sigma^{\prime}=\{\tilde{\uparrow},\tilde
{\downarrow}\}$, as functions of the injection energy $E$. Obviously, for any
$\left\vert \alpha\right\vert =\left\vert \beta\right\vert $ with $\phi
^{B}=(2n+1)\pi$ and $n\in\operatorname{integer}$, the ring is completely
destructive for electron spins of any directions, namely, $\phi_{\uparrow
\left[  \theta,\phi\right]  }=\phi_{\downarrow\left[  \theta,\phi\right]
}=(2n+1)\pi$. We thus have the insulating state in Fig. \ref{fig3}(a). For the
spin-filtering state in Fig. \ref{fig3}(b), we take the values $(\alpha
,\beta)/\alpha^{\ast}=\pm1\times(0,1)$ and $\phi^{B}=(4n+1)\pi/2$ which
contribute to the $\tilde{\uparrow}$ and $\tilde{\downarrow}$ spins with the
constructive $\phi_{\tilde{\uparrow}}=2n\pi$ and destructive $\phi
_{\tilde{\downarrow}}=(2n+1)\pi$ phases. As a response to the anti-symmetric
property of the $\phi_{\tilde{\uparrow}(\tilde{\downarrow})}$ mentioned above,
the indices $\tilde{\uparrow}$ and $\tilde{\downarrow}$ in the transmission
coefficients are swapped if we interchange $\alpha$ and $\beta$, i.e., choose
the values $(\alpha,\beta)/\alpha^{\ast}=\pm1\times(1,0)$ as used in Ref. \onlinecite{Non-Abelian}.

Here the destructive interference means that the out-going channel is blocked.
So far we have not utilized any relation between the incoming and outgoing
spin states, which are correlated with each other by the spin precession, for
example, the PSH, which yields the spin-flipping ring in Fig. \ref{fig3}(c)
and spin-keeping ring in Fig. \ref{fig3}(d). We now focus on the case of
$\left\vert \alpha\right\vert =\left\vert \beta\right\vert $. Without loss of
generality, assume $\alpha>0$ for simplicity. Due to $U_{\text{phase}%
}(\left\vert \alpha\right\vert \rightarrow\left\vert \beta\right\vert )\propto
I_{s}$, the phases are now merely determined by the magnetic gauge and become
spin-independent, i.e. $\phi^{B}=\phi_{\uparrow_{\left[  \theta,\phi\right]
}}=\phi_{\downarrow_{\left[  \theta,\phi\right]  }}$. As a result, the
interference and the precession are decoupled in the PSH, and are governed
separately by the magnetic and the SO gauges.

In one of the equal-strength case, namely, $\alpha=\beta$, we have, according
to Eq. (\ref{Ua=b}), the PSH pattern shown in Fig. \ref{fig1}(a). The
condition $\sqrt{2}l=L_{P}(2s+1)/2$, with $L_{P}$ defined in Eq. (\ref{Lp}),
or equivalently
\begin{equation}
\alpha=\beta=\frac{\left(  2s+1\right)  \pi\hbar^{2}}{4\sqrt{2}lm}%
\text{,}\label{aflipcond}%
\end{equation}
rotate any injected spin $\uparrow_{\left[  \theta,\pi/4\right]  }$ (or
$\downarrow_{\left[  \theta,\pi/4\right]  }$) along the $x=y$ plane, by the
angle $\theta_{\text{PSH}}^{+}=\pi(2s+1)$. For example, the sketch in Fig.
\ref{fig1}(a) is the case with $\theta_{\text{PSH}}^{+}=$ $3\pi$. As a result,
after passing through the ring, $\uparrow_{\left[  \theta,\pi/4\right]  }$ is
flipped to $\downarrow_{\left[  \theta,\pi/4\right]  }$ (or $\downarrow
_{\left[  \theta,\pi/4\right]  }$ to $\uparrow_{\left[  \theta,\pi/4\right]
}$). This represents the spin-flipping ring. Indeed, under the condition
(\ref{aflipcond}), and with $B=0$, our numerical result in Fig. \ref{fig3}(c)
suggests the spin-flipping state by showing vanishing $T_{\uparrow_{\left[
\theta,\pi/4\right]  }\uparrow_{\left[  \theta,\pi/4\right]  }}$ and
$T_{\downarrow_{\left[  \theta,\pi/4\right]  }\downarrow_{\left[  \theta
,\pi/4\right]  }}$ but non-vanishing $T_{\uparrow_{\left[  \theta
,\pi/4\right]  }\downarrow_{\left[  \theta,\pi/4\right]  }}$ and
$T_{\downarrow_{\left[  \theta,\pi/4\right]  }\uparrow_{\left[  \theta
,\pi/4\right]  }}$.%

In the other equal-strength case, $\alpha=-\beta$, we have, according to Eq.
(\ref{Ua=-b}), the PSH pattern shown in Fig. \ref{fig1}(b). Therefore, as long
as the ring retains its square shape, injected spin of arbitrary direction
remains in its original configuration due to the zero precession angle
$\theta_{\text{PSH}}^{-}=0$. One thus arrives at the spin-keeping ring in
accord with Fig. \ref{fig3}(d) which predicts such a state by showing
vanishing $T_{\uparrow_{\left[  \theta,\phi\right]  }\downarrow_{\left[
\theta,\phi\right]  }}$ and $T_{\uparrow_{\left[  \theta,\phi\right]
}\downarrow_{\left[  \theta,\phi\right]  }}$. Note that although Figs.
\ref{fig3}(c) and \ref{fig3}(d) are plotted for the special case $B=0$, the
features of spin-flipping and spin-keeping are robust against magnetic field
since they result from precession rather than phase change.

\section{Summary\label{Summary}}

In the limit Eq. (\ref{FDcond}), the approximation Eq. (\ref{AxAyapp})
suggests that the \textit{SU}(2) or SO phases can be treated as \textit{U}(1)
phases. We have justified the SO-interacting TB model, established from this
analogy, with the one previously given in Ref. \onlinecite{Nicolic}. The PSH,
initially obtained from a global \textit{SU}(2) transformation,\cite{PSHSCZ}
is revisited here by performing a local gauge transformation. As an
application of the geometry of the PSH pattern, we consider a square ring in
contact with two ideal leads. This setup is found to be a versatile
spintronics device performing four types of functions in which spins are
insulated, filtered, flipped or kept (in polarization). The former two are due
to phase change, while the latter two are due to spin precession and newly
proposed here. The SO phases in the presence of both Rashba and Dresselhaus SO
couplings are also analyzed. In particular, the SO phases, as functions of the
two SO couplings, exhibit high symmetries and show discontinuities. Our
numerical results on transmission coefficients confirm with theoretical
predictions on the PSH and thus suggest a four state device.

\begin{acknowledgments}
One of the authors (S.H.C.) thanks Ming-Hao Liu for valuable discussions and
suggestions. This work is supported by the Republic of China National Science
Council Grant No. 95-2112-M-002-044-MY3.
\end{acknowledgments}


\begin{thebibliography}{99}
\bibitem {Footnote}This symmetry yields $T_{\sigma^{\prime}\sigma}$ unaffected
when replacing $\alpha$, in Hamiltonian (\ref{RDha}) with $-\alpha$, as used
in Ref. \onlinecite{Non-Abelian} where $\beta$ is further dropped.
\end{thebibliography}
\end{document}